\documentclass[letters,usenatbib]{mnras}

\usepackage{newtxtext,newtxmath}
\usepackage{txfonts}

\usepackage[T1]{fontenc}
\usepackage{ae,aecompl}

\usepackage{graphicx}	
\usepackage{amsmath}	
\usepackage{amssymb}	

\newcommand{\kms}{\mbox{${\rm km\,s}^{-1}$}}
\newcommand{\mjyb}{\mbox{${\rm mJy\,beam}^{-1}$}}
\newcommand{\doce}{\mbox{$^{12}{\rm CO}$}}
\newcommand{\trece}{\mbox{$^{13}{\rm CO}$}}
\newcommand{\dieciocho}{\mbox{${\rm C}^{18}{\rm O}$}}

\title[A ringed concentration of mm-grains in Sz\,91]{A ring-like concentration of mm-sized particles in Sz\,91}

\author[H. Canovas et al.]{
H. Canovas,$^{1,2}$\thanks{E-mail: hector.canovas@dfa.uv.cl}
C. Caceres,$^{1,2}$
M.~R. Schreiber,$^{1,2}$
A. Hardy$^{1,2},$
L. Cieza$^{2,3},$\newauthor
F. M\'enard$^{4},$
A. Hales$^{5,6}$\\
$^{1}$Departamento de F\'isica y Astronom\'ia, Universidad de Valpara\'iso, Valpara\'iso 2360102, Chile\\
$^{2}$Millennium Nucleus ``Protoplanetary discs in ALMA Early Science", Universidad de Valpara\'iso, Valpara\'iso 2360102, Chile\\
$^{3}$Facultad de Ingeniería, Universidad Diego Portales, Av. Ejercito 441, Santiago, Chile\\
$^{4}$UMI-FCA, CNRS/INSU, France (UMI 3386), and Departamento de Astronomía, Universidad de Chile, Casilla 36-D Santiago, Chile\\
$^{5}$Atacama Large Millimeter/Submillimeter Array, Joint ALMA Observatory, Alonso de C\'ordova 3107, Vitacura 763-0355, Santiago - Chile\\
$^{6}$National Radio Astronomy Observatory, 520 Edgemont Road, Charlottesville, VA, 22903-2475, USA
}

\date{Accepted 2016 January 6. Received 2015 December 20; in original form 2015 November 16}

\pubyear{2015}

\begin{document}
\label{firstpage}
\pagerange{\pageref{firstpage}--\pageref{lastpage}}
\maketitle

\begin{abstract}
Models of planet formation and disc evolution predict a variety of observables in the dust structure of protoplanetary discs. 
Here we present Atacama Large Millimeter/submillimeter Array (ALMA) Band-6 and Band-7 observations of the transition
disc Sz\,91 showing that the continuum emission at 870$\mu$m, which is dominated by emission from large dust grains,
is localized in an optically thin narrow ring.
We find that most of the emission ($\sim95\%$) is concentrated in a ring located at 110 au from the central star that is only
about 44 au wide. In contrast, the $^{12}\mathrm{CO}$ (2-1) emission peaks closer to the star and is detected up to $\sim488$
au from the star. The concentration of large grains in a ring-like structure while the gas disc extends much further in and
further out is in qualitative agreement with predictions of hydrodynamical models of planet-disc interactions including radial
drift and gas drag.

\end{abstract}

\begin{keywords}
protoplanetary discs -- stars: variables: T Tauri -- planet-disc interactions -- stars: individual (Sz\,91)
\end{keywords}

\section{Introduction} \label{sec:intro}
The direct detection of giant planets in protoplanetary and debris discs 
\citep[]{2010Sci...329...57L, 2013ApJ...766L...1Q,2015Natur.527..342S}
confirms that planets are born in protoplanetary discs. The detailed planet
formation process, however, is rather poorly understood. For a long time, dust
grains were expected to drift inwards on time-scales too fast  to allow for significant
growth of solids \citep{1977MNRAS.180...57W}. However, recent simulations
suggest that radial-drift is not as effective as expected, showing that grains
could survive the so called `radial-drift barrier' \citep{2012A&A...537A..61L, 2014MNRAS.437.3037L}.
High spatial resolution observations of potentially planet forming discs that allow
the measurement of the radial distribution of dust grains are crucial to constrain
the importance of radial drift for the planet formation process.

Models predict that a forming giant planet should carve a gap/cavity in its host
protoplanetary disc, causing a pressure maximum at the inner edge of the disc.
This has dramatic consequences on the distribution of  dust particles. The larger
particles in the outer disc drift inwards until they are held back near the pressure
maximum \citep{2006A&A...453.1129P, 2007A&A...474.1037F}. On the
other hand, small particles ($\lesssim10\mu$m) do not accumulate at the pressure
maximum and are allowed to enter the cavity \citep[e.g.][]{2006MNRAS.373.1619R}.
Thus, if a giant planet has carved a gap or cavity inside a protoplanetary disc, an
accumulation of large particles in a ring-like structure close to the inner  edge of
the outer disc is expected \citep[][]{2012A&A...545A..81P}. The sizes of the
particles affected by this filtering/trapping mechanism depend on the exact shape
of the pressure bump which mostly depends on the planet mass \citep{2012A&A...547A..58G, 2013A&A...560A.111D}. 
The two key ingredients of this scenario, dust filtration and radial drift, are supported by
recent observations. The different cavity sizes found for different grain sizes in several
transition discs confirm dust filtration \citep[e.g.,][]{2013A&A...560A.105G}. Furthermore,
radial drift triggered by gas drag \citep{2014ApJ...780..153B} can explain the observed
difference in the outer radii of the gas and dust components of protoplanetary discs
\citep[e.g., ][]{2009A&A...501..269P, 2012ApJ...744..162A, 2013A&A...557A.133D}.

In this work we present ALMA cycle\,2 data of the transition disc Sz\,91. This young system
\citep[age $\sim1$Myr, $M_\star = 0.47\,\mathrm{M_{\sun}}$, $T_{\star} = 3720$ K, see]
[hereafter C2015, for details]{2015ApJ...805...21C} is located in the Lupus III star forming
region at a distance of 200\,pc \citep{2008ApJS..177..551M}, and was first classified as a
potentially planet forming disc by \citet{2012ApJ...749...79R} based on its SED. Further
observations revealed that Sz 91 has remarkably different cavity sizes as a function of
wavelength: $\sim65$ au in radius at $K_\mathrm{s}$ band \citep[][]{2014ApJ...783...90T}, which is
sensitive to small grains, and $\sim97$ au at 1.3 mm (C2015), which traces large grains. 
The outer edge of the gaseous disc has also been detected further than the continuum
emission at 1.3 mm (C2015). However, due to the resolution and sensitivity of these
observations, it was previously not possible to assess whether that discrepancy was due
to a physical difference between the radial distribution of gas and large dust, or due to
an observational effect. The accretion rate ($7.9\times10^{-11} \mathrm{M yr^{-1}}$) and
huge cavity exclude photoevaporation as the mechanism responsible for the gap formation
and the lack of a detected stellar companion makes Sz\,91 one of the strongest candidates
for a planet forming disc (C2015). Our new observations further support this hypothesis.
The new data shows that large grains are distributed in a narrow, ring-like structure as
predicted by  models of planet-disc interactions, while $^{12}\mathrm{CO}$ (2-1) is
detected much further out from the central star. This accumulation of dust may facilitate
the growth of solids.

\section{Observations} \label{sec:obs}

Sz\,91 was observed with ALMA in band-6 and band-7 with two different configurations
during Cycle-2 (programme 2013.1.00663.S, PI: Canovas).

In Band 6, 40 (12-m) antennas were used during the observations with baselines ranging
from 23.3 to 558.2 m (17.9 to 429.4 $\mathrm{k\lambda}$). The ALMA correlator was
configured to provide one continuum spectral window with a total bandwidth of 2 GHz
centred at 231.6 GHz and three spectral windows with bandwidths of 58.6 MHz and 
channel widths of 61.035 KHz. These narrow windows were centred on the $\doce$
(2-1, 230.5 GHz) (hereafter $\doce$), $\trece$ (2-1, 220.4 GHz), and $\dieciocho$
(2-1, 230.5 GHz) lines. 

The Band 7 observations were performed with 42 (12-m) antennas with baselines ranging
from 15.1 to 1574.4 m (17.3 to 1789.1 $\mathrm{k\lambda}$). The correlator was configured
to provide two continuum spectral windows with a total bandwidth of 2 GHz centred at 332.5
and 343.7 GHz, and two spectral windows with bandwidths of 58.6 MHz and channel widths
of 61.035 KHz aimed to measure the $^{12}$CO (3-2, 345.8 GHz) and $^{13}$CO (3-2, 330.6 GHz).

The quasars QSO J1517-2422 and QSO J1610-3958 were observed for bandpass and phase
calibration. Titan and Ceres were observed to perform flux calibration in Band 6 and Band 7, respectively.
The uncertainty associated with the flux calibration is $\sim10\%$ (ALMA Cycle 3 Technical Handbook). 
Water vapor was continuously monitored with the radiometers attached to each antenna to correct for
the fast fluctuations of the phase. In both bands the individual exposure times were 6.05 s,
amounting to a total exposure time for the science observations of 207 and 1430 s in Band 6
and Band 7, respectively.  The median precipitable water vapor (PWV) was 0.67 mm and 1.15 for
Band 6 and Band 7, respectively.

In this letter we focus on the observational constraints derived from the Band 7 continuum and the
$\doce$ moment maps, as they provide the best trade-off between spatial resolution and sensitivity.
A detailed model of the dusty and gaseous disc is beyond the scope of this letter and will be presented
in a forthcoming paper.

We used CASA \citep[][]{2007ASPC..376..127M} to process the visibilities. Channels showing artefacts
were flagged out. The line-free channels were used to fit and subtract the continuum to the emission-line
bearing channels. The CLEAN algorithm \citep{1974A&AS...15..417H} was applied to deconvolve
the visibilities using Briggs and natural weighting for the $\doce$ and continuum, respectively. We
applied one round of self-calibration finding a marginal increment in S/N, so no more iterations were
applied. The median rms in the continuum is 0.95 $\mjyb$ with a beam size of 0.21 arcsec $\times$ 0.15 arcsec
($42\times30$ au) and position angle (PA) of $79.9\degr$. The beam size in the $\doce$ data is
0.60 arcsec $\times$ 0.57 arcsec ($120\times114$ au) and PA of $84.2\degr$. The $\doce$ is detected
above $3\sigma$ at velocities ranging from 0.320 to 5.040 $\kms$ (LSRK), with a median
rms in the line-free channels of $20.12$ $\mjyb$. 

\begin{figure*}
\center
\includegraphics[width = 0.85\linewidth,trim = 0 20 0 0]{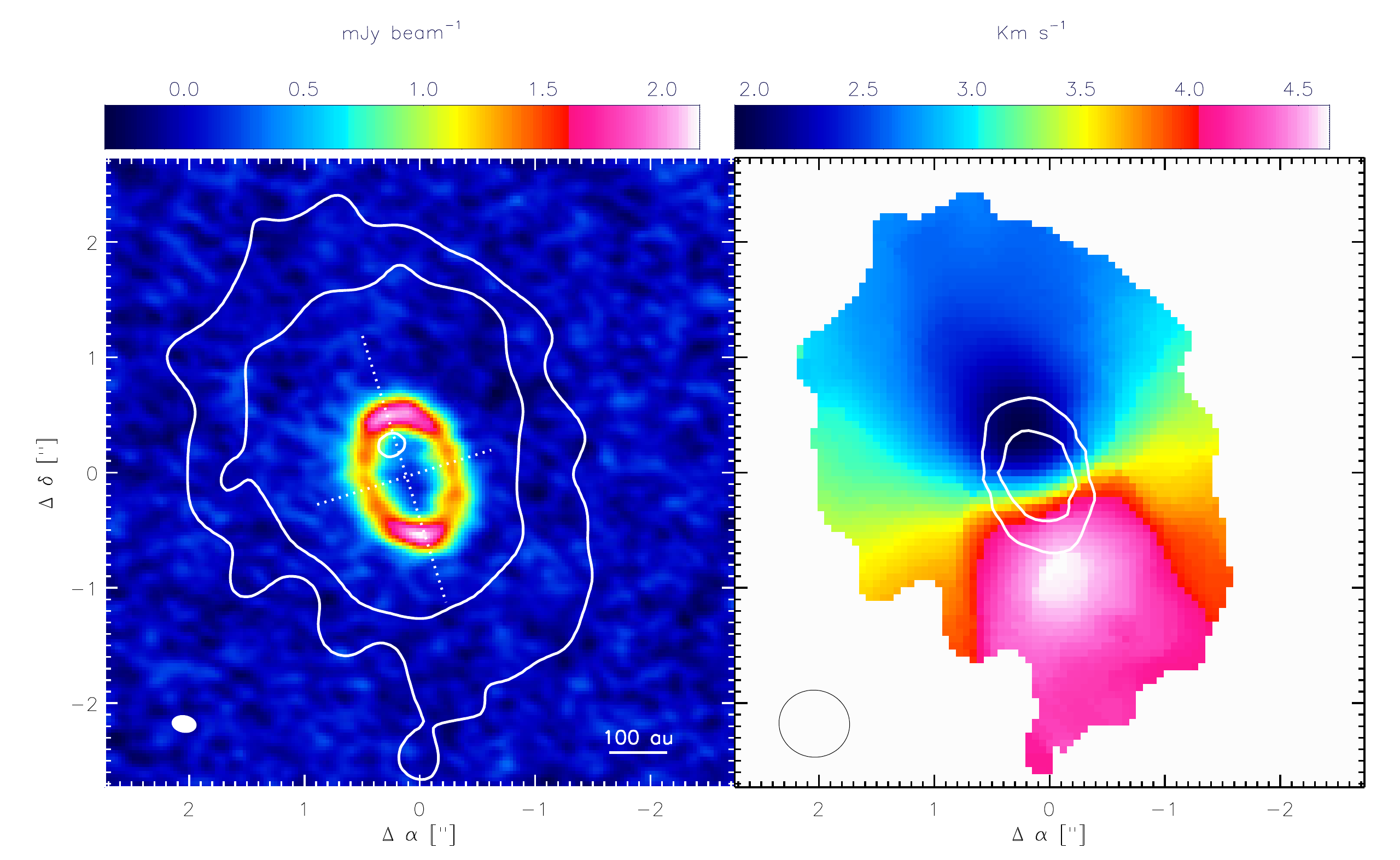}
\caption{\textbf{Left:} cleaned continuum image at band-7 (870$\mu$m). The dotted cross shows the
disc's centre and major/minor axis derived in Sect.~\ref{sec:sub_results_dust}. White lines contour the
$\doce$ integrated intensity (Moment-0) map at $(2.5,15,31)\times$ rms ($22.7 \mjyb$$\kms$), highlighting
that the $\doce$ emission is detected much further out than the continuum emission. The $\doce$ emission
peaks inside the ring's hole, at $\sim60$ au. The asymmetric profile is a consequence of the cloud
contamination (see text). \textbf{Right:} $\doce$ velocity field (Moment-1) map. The white lines contour the
$10\times$ rms (0.1 mJy) of the continuum emission. The synthesized beam of the continuum and $\doce$
observations is shown in the bottom-left corner of the left and right figures, respectively.}
\label{fig:fig1}
\end{figure*}

\section{RESULTS} \label{sec:results}

\subsection{Continuum} \label{sec:sub_results_dust}
The continuum image (Fig.~\ref{fig:fig1}, left) shows an inclined narrow ring with a large central
hole clearly resolved. The peak flux is $2.17\pm0.95$ $\mjyb$. Integrating the flux inside the
regions with S/N $>3\sigma$ results in $\mathrm{F_{1.3\mathrm{mm}}} = 40.15 \pm 0.95$ mJy. 
We derive the geometry of the continuum disc by fitting a ring profile to the continuum visibilities
using \textit{uvfit}/{\sc{MIRIAD}}. We find that the disc's centre is offset by
$\Delta\alpha = $ 0.10 arcsec $\pm$ 0.004 arcsec and $\Delta\delta =$ -0.02 arcsec $\pm$ 0.004
arcsec from the ALMA phase centre ($\alpha_{2000} = 16^{\mathrm{h}}07^{\mathrm{m}}11^{\mathrm{s}}.56$
and $\delta_{2000} = -39{\degr}03\farcm47\farcs82$). We derive a PA = 17.4$\pm0.4\degr$, inclination
$i = 51.7\pm0.4\degr$, and a total flux of $42.43 \pm 0.95$ mJy. This value is slightly higher than
previously reported observations at the same wavelength \citep{2012ApJ...749...79R, 2014ApJ...783...90T},
and we consider it very likely that this difference is a consequence of calibration uncertainties.

The hole of the ring shows no emission above the rms of our images. The emission at the
ring's ansae is higher than in the rest of the ring. This can be naturally explained by a combination
of the disc's inclination and optically thin emission, as there is more dust emitting along the line of
sight of the major axis.

The dust mass of the ring can be roughly estimated using the linear relations derived by
\citet{2005ApJ...631.1134A, 2007ApJ...671.1800A} and calibrated by \citet{2008ApJ...686L.115C}:
\begin{eqnarray}\label{eq:mass}
M_{\rm{dust}} \sim8.0 \times 10^{-5}~[\frac{F_{\nu} (0.86~\rm{mm})} {\rm{mJy}} \times (\frac{d}{140~\rm{pc}})^2]~M_{\rm{Jup}},
\end{eqnarray}
where $d$ is the distance to the disc. Using a distance of 200 pc \citep{2008ApJS..177..551M}, the
measured total flux, and $10\%$ uncertainties for the flux and distance, we derive a total mass of
$23\pm6\,M_\mathrm{Earth}$, in agreement with previous estimates for the disc's mass in large
grains \citep{2012ApJ...749...79R, 2014ApJ...783...90T, 2015ApJ...805...21C}. However we are
cautious about this result as a major assumption in this method is that the grain size distribution
follows an interstellar medium (ISM) distribution with power law index $p = -3.5$ \citep{1977ApJ...217..425M},
and this may not be the case when the dust population is dominated by large grains. A detail model
exploring different possible scenarios is needed to better constrain the mass in the ring but this is
beyond the scope of this letter.

\subsection{$\doce$ Moments} \label{sec:sub_results_gas}

In order to evaluate the total extension of the gaseous disc, we created the Moment-0
(velocity-integrated intensities) and Moment-1 (intensity-weighted velocities) images
from the channels showing significant ($>3\sigma$) emission. Pixels with values below
$2.5\times$ the median rms ($20.12$ $\mjyb$) were excluded from these computations. 
A Gaussian profile was fitted along the spectral axis at each spatial position to ensure
that the full $\doce$ profile was included at each location. The Moment-0 has an rms
of 22.7 $\mjyb \kms$ and its integrated flux is $4.932\pm0.023$ Jy $\kms$. Fig.~\ref{fig:fig1}
(right) displays the Moment-1 in colours encompassing the $2.5\times$rms region
of the Moment-0. The asymmetric shape of the images is a consequence of the filtering-out of the
interferometer due to cloud emission at $v_\mathrm{LSRK}>3.8$ $\kms$, which
translates to a reduced flux in the red-shifted side of the $\doce$ line \citep[previously
noticed by][]{2014ApJ...783...90T, 2015ApJ...805...21C}. The blue-shifted side, unaffected by the cloud,
shows $\doce$ emission up to $\sim2.44$ arcsec ($\sim$488 au) from the disc's centre,
and peaks at $60\pm12$ au, inside the hole of the ring and near the cavity edge
observed at $K_\mathrm{s}$ band \citep{{2014ApJ...783...90T}}.

\begin{figure}
\center
\includegraphics[width = 1.0\linewidth,trim = 20 40 20 10]{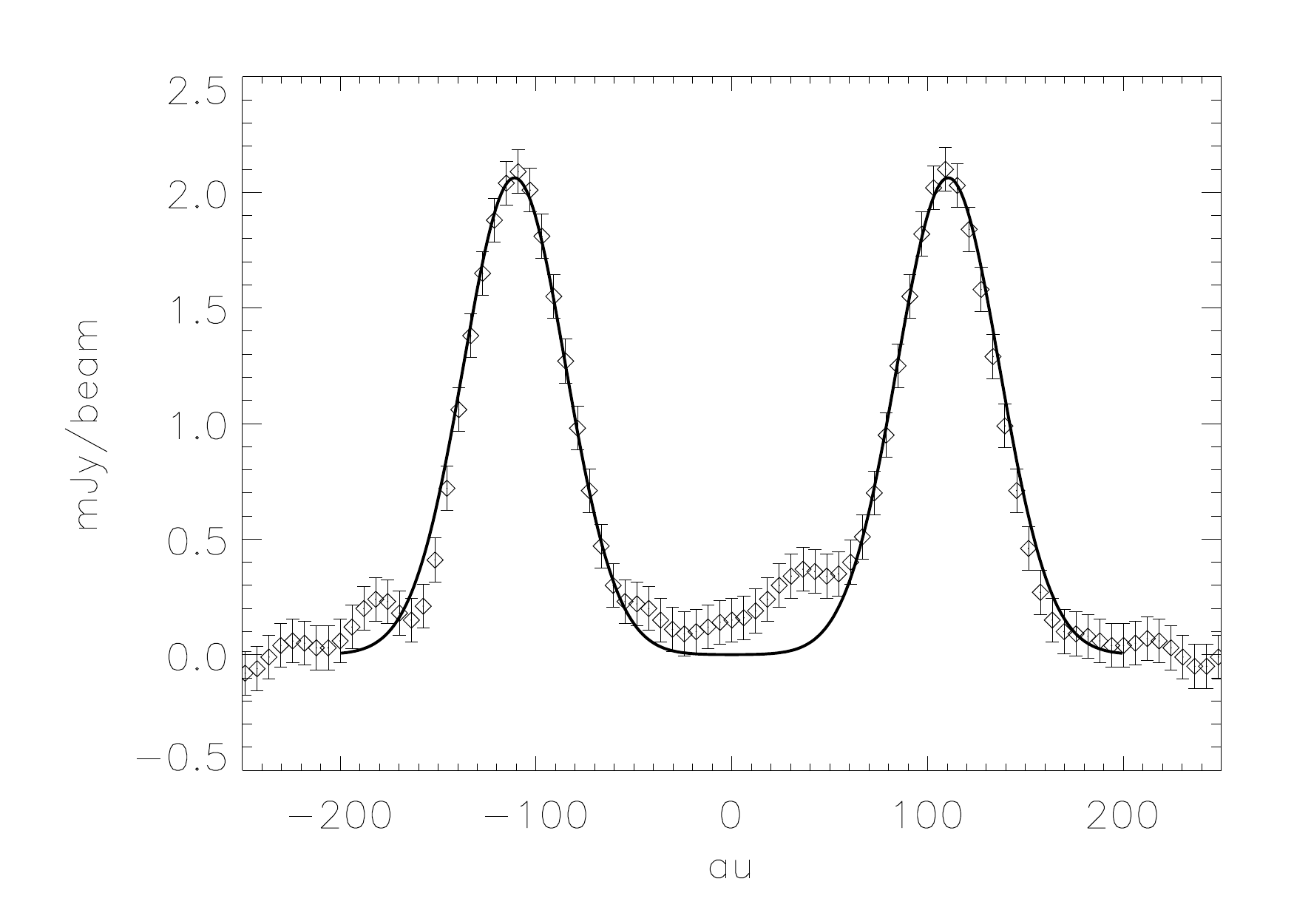}
\caption{Cut along major axis of the Band-7 continuum disc. Error bars represent the rms of the observations;
the peak is detected with a signal to noise of S/N$\sim$22. Positive distance points to the north direction,
and negative distance to the south direction. The ring has same width and peak values along the northern
and southern sides. Inside the hole, there is a small excess ($\lesssim3\sigma$) of emission along the
northern side when compared to its southern counterpart. The solid curve represents the best fit described
in Sect.~\ref{sec:sub_constrains_dust}.}
\label{fig:fig2}
\end{figure}

\subsection{Surface brightness profile of mm emission} \label{sec:sub_constrains_dust}

To derive further constraints on the ring geometry we use the brightness profile along the major axis of the
ring. We find that the emission is concentrated in a narrow ring between $\sim60$ and $\sim150$ au from
the central star, with the peak emission at $\sim110$ au (Fig.~\ref{fig:fig2}).  Averaging the emission in ellipsoids
which have the same inclination and PA as the disc provides the same results. We then fit three
different brightness profiles to the surface brightness profile along the major axis: a Gaussian, a constant (hat-like),
and a power-law with exponent $-1.0$. We vary the widths and central positions of these profiles, and convolve
them with a Gaussian function of full width at half-maximum FWHM =  0.16 arcsec (i.e., equal to the FWHM of the
continuum synthesized beam along the disc's major axis) to compare with the observations. We then perform a
Monte Carlo Markov Chain (MCMC) exploration to identify the best fit to the observations. 
We find that a Gaussian profile centred at $110.5\pm0.3$ au and with $1\sigma$ width of $22.2\pm0.5$
au results in the best fit to our observations, with reduced $\chi_{\nu}^2 = 1.27$ (Fig.~\ref{fig:fig2}). For comparison,
the hat-like and the power-law distribution result in $\chi_{\nu}^2 = 1.61$ and  $\chi_{\nu}^2 = 2.39$, respectively.
Our best fit implies that $95.5\%$ of the dust probed by our Band-7 observations is located within $88-132$ au.

\section{Discussion} \label{sec:discussion}

The ALMA Cycle-2 observations presented here have much higher sensitivity and spatial 
resolution than previous observations. They reveal two striking features: (1) there is a large
difference between the outer disc radius of the large grains and the $\doce$ disc, and (2) the
large grains are concentrated in a narrow, ring-like structure. The difference in the outer edges
of the gaseous and continuum discs cannot be explained by beam dilution or sensitivity effects,
and overall our observations show that the radial distribution of the large particles probed by
the continuum images is very different from the radial distribution of the gas.


\subsection{Sz\,91 radial structure} \label{sec:sub_radial_dust}
The spectral energy distribution (SED) of Sz\,91 indicates that the innermost regions of the disc must contain
a very small amount of dust to simultaneously reproduce the excess emission at $12\mu$m and the absence
of emission above photospheric levels at wavelengths below $10\mu$m \citep[][]{2012ApJ...749...79R, 2014ApJ...783...90T, 2015ApJ...805...21C}.
Imaging polarimetry observations with SUBARU at $K_\mathrm{s}$ band resolved the disc showing that Sz\,91
has a $\sim65$ au cavity in its $\mu$m-sized grain distribution \citep{2014ApJ...783...90T}
while ALMA Band-6 (1.3 mm) Cycle-0 observations revealed that the radius of the cavity is significantly
larger for mm-sized particles, i.e. $\sim97$ au (C2015). Combining these observational constraints and
the observations presented in this work, and taking into account that the small grains are well coupled to
the gas \citep{2007A&A...474.1037F}, a complex radial structure emerges. The distribution of small
grains is severely depleted inside the inner $\sim65$ au, and extends beyond 400 au. On the other hand,
the large grains that contain the bulk of the disc mass in dust are concentrated in a ring. Most of the emission
is located between $\sim88$ and $\sim132$ au. As shown by C2015, $\doce$ is detected down to $\sim28$
au from the star, well inside the ring's hole.

\subsection{A planet-induced pile up?} \label{sec:sub_mechanisms}

Several observations of transition \citep[e.g.,][]{2012ApJ...744..162A, 2013ApJ...775..136R, 2014ApJ...791L...6W}
and protoplanetary discs \citep[e.g.][]{2009A&A...501..269P, 2013A&A...557A.133D} show that the gaseous
disc extends further out than the dusty disc. In all these cases, radial drift of large particles \citep{2014ApJ...780..153B}
has been invoked as the natural explanation for these discrepancies. In the case of Sz\,91, the differences in the
outer edges of the $\doce$ disc and the geometry of the distribution of large particles (inner and outer edges), indicates
that there must be at least one other mechanism in action. The effect of multiple giant planets orbiting inside the cavity
seems plausible, given the observed characteristics of this object (see C2015 for a detailed discussion). 
The different cavity sizes for the small and large grain distributions are in agreement with the predictions of recent
hydrodynamical simulations of multiple planets embedded in a disc \citep[][]{2015ApJ...809...93D, 2015A&A...573A...9P}. 
Hydrodynamical models describing the planet-disc interactions including radial drift predict that the large grains 
will concentrate at the pressure maximum created by an orbiting planet, resulting in a narrow, ring-like structure
\citep{2012A&A...547A..58G, 2012A&A...545A..81P}. Our observations are in agreement with these predictions.

While it is clear that the mm sized particles are located in a narrow ring, our observations are unable to tell if the
observed structure represents a stationary state. If the torque created by the orbiting planets equals the drift torque
this would be the case but it might also be that the large grains are currently still in the process of piling up.

\section{Conclusions} \label{sec:sub_cavity}

In this letter we present ALMA Cycle 2 observations of the continuum at $870\mu$m and $\doce$
of the transition disc Sz\,91. The images reveal that the distribution of the gas and the large grains
is very different, with the large particles detected up to $\sim150$ au while  the gas is detected up to
$\sim488$ au. We find that a Gaussian profile centred at $110.5\pm0.3$ au and with $1\sigma$
width of $22.2\pm0.5$ au provides the best fit to our observations, which means that most of the
continuum emission is located in a narrow ring between $\sim88-132$ au.
Our findings are in qualitative agreement with predictions of a pile-up of large grains due to a
combination of pressure bumps created by low-mass companions, and radial drift triggered by
gas drag \citep{2012A&A...545A..81P,2013A&A...560A.111D}.  This concentration of
large grains could facilitate the formation process of further planets as the particles may grow
faster because of enhanced density of solids.

The hole at (sub)mm wavelengths is surprisingly large when compared with transition discs around low-mass
stars \citep[e.g., ][]{2011ApJ...732...42A} which, together with its moderate accretion rate, suggests the presence
of multiple planets inside the hole \citep{2011ApJ...738..131D}. The presence of a small amount of $\mu$m-sized
grains in the innermost regions of the disc, the difference in cavity radii between the $\mu$m grains and  mm-sized
grains, and the pile-up of large grains in a narrow ring, are consistent with a giant planet orbiting at the inner edge
of the $\sim65$ au cavity observed at $K_\mathrm{s}$ band \citep{2013A&A...560A.111D}. Given the wealth of signposts
of planet-disc interactions in Sz\,91, we consider this object an ideal laboratory with which to test model predictions
for planet formation, planet disc interaction, and protoplanetary disc evolution around low-mass stars.

\section*{Acknowledgements}
We thank the referee for his/her useful comments and suggestions.
HC acknowledges support from ALMA/CONICYT (grants 31100025 and 31130027).
CC and MRS acknowledge support from FONDECYT grants 3140592 and 1141269. AH is supported
by Doctoral scholarship FIB-UV 2014 and Gemini/Conicyt 32120033 LC and FM acknowledge funding
from CONICYT FONDECYT \#1140109 and the EU FP7-2011
(Grant Agreement no. 284405), respectively. This paper makes use of the following ALMA data:
ADS/JAO.ALMA\#2011.0.00733.S. ALMA is a partnership of ESO (representing its member states),
NSF (USA) and NINS (Japan), together with NRC (Canada), NSC and ASIAA (Taiwan), and KASI
(Republic of Korea), in cooperation with the Republic of Chile. The Joint ALMA Observatory is
operated by ESO, AUI/NRAO and NAOJ. The National Radio Astronomy Observatory is a facility
of the National Science Foundation operated under cooperative agreement by Associated Universities, Inc.







\bsp	
\label{lastpage}
\end{document}